# High-precision and low-noise dielectric tensor tomography using a micro-electromechanical system mirror

Juheon Lee,[1,2‡] Byung Gyu Chae,[3‡] Hyuneui Kim,[3] MinSung Yoon,[3] Herve Hugonnet,[1,2] YongKeun Park[1,2,4,*]

[1] Department of Physics, Korea Advanced Institute of Science and Technology (KAIST), Daejeon 34141, Republic of Korea

[2] KAIST Institute for Health Science and Technology, KAIST, Daejeon 34141, Republic of Korea

[3] Holographic Contents Research Laboratory, Electronics and Telecommunications Research Institute, Daejeon 34129, Republic of Korea

[4] Tomocube Inc., Daejeon 34109, Republic of Korea

* yk.park@kaist.ac.kr

**ABSTRACT:** Dielectric tensor tomography is an imaging technique for mapping three-dimensional distributions of dielectric properties in transparent materials. This work introduces an enhanced illumination strategy employing a micro-electromechanical system mirror to achieve high precision and reduced noise in imaging. This illumination approach allows for precise manipulation of light, significantly improving the accuracy of angle control and minimizing diffraction noise compared to traditional beam steering approaches. Our experiments have successfully reconstructed the dielectric properties of liquid crystal droplets, which are known for their anisotropic structures, while demonstrating a notable reduction in background noise of the images. Additionally, the technique has been applied to more complex samples, revealing its capability to achieve a high signal-to-noise ratio. This development represents a significant step forward in the field of birefringence imaging, offering a powerful tool for detailed study of materials with anisotropic properties.

**Table of Contents**

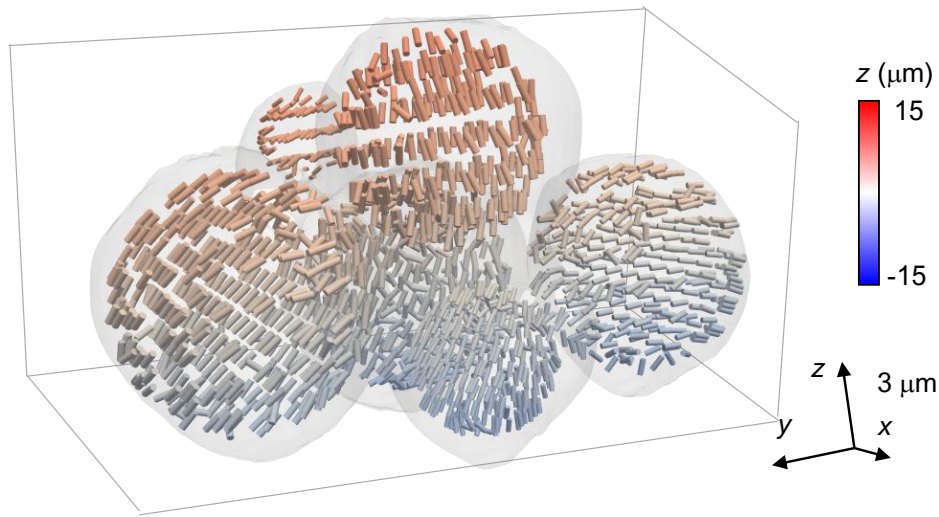

## Introduction

Holotomography (HT) is a three-dimensional (3D) quantitative phase imaging (QPI) technique that reconstructs the three-dimensional refractive index (RI) distribution of transparent samples[1-4], highlighting its crucial role in non-invasive label-free biological imaging across various fields, including immunology,[5-7] cell biology[8-10], neuroscience,[11, 12] and cancer research.[13-15] The advent of dielectric tensor tomography (DTT), an evolved form of HT capable of assessing 3D optical anisotropy, marks a significant expansion of HT's application scope into material science and soft matter physics[16]. DTT reconstructs a sample's dielectric tensor tomogram, offering a comprehensive analysis of light-matter interaction by accounting for light polarization and the orientation and birefringence of molecules.

In DTT, precise control over the illumination angle is paramount, as each optical field, corresponding to a specific angle, must be accurately mapped onto the correct Ewald surface in the 3D Fourier space[17]. Additionally, to capture the complete dielectric tensor elements, it's essential to acquire multiple optical field images at slightly varied illumination angles[16]. Galvanometer-based scanning mirrors, which electromechanically rotate to direct the incident light, are commonly used for angle control (Fig. 1a)[18]. However, their mechanical instability, resulting from electric noise, inertia, and size, poses significant challenges[19]. Moreover, aligning the two independent rotation axes with the image plane simultaneously is difficult, and employing a relay-lens system to circumvent this only leads to a bulkier, less stable optical setup.

Alternative angle control devices, such as spatial light modulators (SLMs) and Digital Micromirror Devices (DMDs), offer certain advantages and limitations. [20, 21] SLMs, leveraging liquid crystals, ensure high stability but suffer from slow response times and high costs, limiting their widespread application. DMDs, on the other hand, use an array of switchable micromirrors to direct the incident beam at specific angles (Fig. 1b). However, the binary nature of DMDs can lead to unwanted diffraction patterns, reducing laser power efficiency and image quality. While techniques like spatial filtering and time-multiplexing can mitigate these issues,[21, 22] they introduce their own set of limitations, including reduced illumination angle range and residual diffraction noise.

This study introduces the use of a micro-electromechanical system (MEMS) mirror for controlling illumination angles in DTT (Fig. 1c). The MEMS mirror, integrating electrical and mechanical components, allows for precise manipulation of a small mirror, maintaining the beam's central position in the image plane. Compared to galvanometer-based mirrors, DMDs, and SLMs, MEMS mirrors offer a simpler implementation and reduced diffraction noise. We evaluate the MEMS mirror's angle precision and accuracy, comparing its performance to that of DMDs in terms of diffraction noise and tomogram background noise. Additionally, we investigate improvements in the signal-to-noise ratio by examining the orientation configuration of complex anisotropic structures.

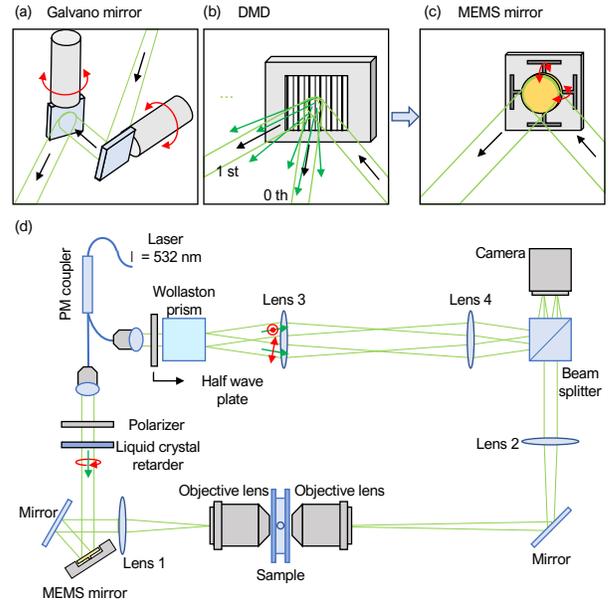

Figure 1. Overview of angle control in DTT setup. (a) Galvanometer scanning mirrors direct light at separate planes. (b) DMD uses a binary pattern for angle-specific light diffraction, with green arrows showing diffraction noise. (c) MEMS mirror reflects light without noise on a single plane. (d) DTT setup schematic incorporating the MEMS mirror.

## Method

The experimental methodology utilizes polarization-sensitive Mach-Zehnder interferometry for capturing the electric field images scattered by a sample. Our setup commences with the division of a laser beam (diode-pumped solid-state laser, wavelength $\lambda$ = 532 nm, LCX-532S, Oxxius laser) into two paths: the sample and reference arms, facilitated by a polarization-maintaining fiber coupler. Polarized illumination is achieved in the sample arm using a linear polarizer coupled with a liquid crystal retarder (LCC1223-A, Thorlabs), enabling the generation of either right or left circularly polarized light. The angles of incident light are precisely controlled via a MEMS mirror (A7M20.2-2000AL, Mirrorcle, with a mirror diameter of 2 mm), supplemented by an additional mirror for direction. The reflected beam is subsequently focused onto the sample by passing through a tube lens (focal length $f$ = 100 mm) and a condenser lens (UPLSAPO 60XW, 60× magnification, numerical aperture NA = 1.2).

The diffracted, polarization-sensitive electric fields are recorded using a spatially multiplexed approach specific to DTT[23]. For the generation of polarization-sensitive interferograms, the reference beam is split into orthogonal polarizations by a Wollaston prism, with a half-wave plate positioned upstream to ensure even beam division. These beams are collected and refocused through a $4f$ telescopic imaging system (both lenses with $f$ = 300 mm) before being recombined with the sample beam via a beam splitter. This configuration ensures that only fields with matching polarization components interfere, producing two distinct polarization-sensitive interferograms. These interferograms are multiplexed onto an image sensor (IMX421, Sony), resulting in fringe patterns that form a checkerboard pattern due to variance in the reference beam angles. Vector fields are extracted from these multiplexed interferograms through Fourier space filtering.

In our DTT experiments, circular scanning was performed with the MEMS mirror across 49 different illumination angles, controlled by modulating voltage signals for the x and y axes. It's important to note that the MEMS mirror, when set to a perpendicular orientation relative to the sample, deviates from the conjugate image plane; thus, voltage signal calibration was necessary to account for the mirror's angled positioning.

For each illumination angle, off-axis holograms were acquired three times, each with a distinct polarization state of the illumination. During the third acquisition, the illumination angle was slightly adjusted from its original setting. The captured sample fields were then leveraged to reconstruct the dielectric tensor tomograms, adhering to DTT principles.[16] Post-reconstruction, the dielectric tensors were diagonalized to determine the principal refractive indices and optic axis orientations, revealing the molecular alignment within the sample.

### Results and discussion

To verify the MEMS mirror's accuracy and precision for DTT, we analyzed the spatial frequencies of the plane waves in the interferograms. The measurements were conducted with one reference beam by eliminating the Wollaston prism from the set-up. Figure 2a shows the recorded fringe patterns along-side with the 2D Fourier transform of the recorded field in the absence of a sample. In the Fourier space, the highest intensity peak represents the main spatial frequency of the incident plane wave.

showing average spatial frequencies (red circles) versus expected values (crosses), with individual frequencies (black circles) and their standard deviations in the inset box plot. (c) Normalized 2D Fourier spaces comparing normal illumination by DMD and MEMS mirror, with condenser NA bandwidth indicated by white dashed arrows. (d) Angular intensity plots derived from the Fourier spaces in (c). (e) Angular intensity plots for circular scanning, with bold lines representing the average for DMD and MEMS mirror.

For statistical analysis, ten circular scans were performed at a speed of 160 Hz, and their spatial frequencies were measured. Subpixel accuracy of the frequencies was achieved by fitting phase ramps to the transmitted phase. Figure 2b shows that the incident spatial frequencies are accurate. However, errors arise due to the nonlinear relation between voltage signals and angles of the MEMS mirror at high angle.[24]

To evaluate precision, we calculated standard deviations of spatial frequencies at each illumination angle (Fig. 2b). The standard deviations are below 0.001 $\mu m^{-1}$. It is noteworthy that the distribution of the standard deviations falls within the acceptable error range of 0.014 $\mu m^{-1}$ needed for slightly tilted illumination in DTT.[16]

To put the SNR in perspective with other angle scanning methods, we compared the normalized 2D Fourier spaces of the fields generated by DMD and MEMS mirror (Figure 2c). The DMD results were obtained using the setup described in the previous study.[23] The Fourier spaces were captured at normal illumination, and cropped within the bandwidth of the objective NA. The white dashed arrows indicate the bandwidth of the condenser NA. In the case of the DMD, time-multiplexed illumination was employed to mitigate undesired diffraction induced by a binary pattern.[22] Despite these efforts, the Fourier space of the DMD exhibits high contrast at the boundary of the condenser NA, indicating undesired diffraction noise collected by the condenser lens. Conversely, the Fourier space of the MEMS mirror exhibits low contrast at the boundary of the condenser NA.

For further visualization of the diffraction noise, we plotted angular intensities for the Fourier spaces in Fig. 2c (Fig. 2d). In Fig. 2e, angular intensities for individual illuminations of circular scanning were also plotted. To mitigate the edge effect in the Fourier spaces, we applied apodization by convoluting the Hanning window to the measured fields.[25] As shown, both DMD and MEMS mirror plots exhibit peak intensity at the spatial frequencies corresponding to the illumination angles. However, the angular intensity of the DMD within the condenser NA is higher than that of the MEMS mirror because of diffraction noise. This noise can decrease the SNR at high frequencies, consequently deteriorating the quality of tomograms as later shown in Figure 3.

To validate the SNR improvement, we compared background noise in 3D dielectric tensor tomograms obtained using the DMD and the MEMS mirror. The DMD results were obtained using the setup described in the previous study.[23] The measurements were performed on liquid crystal (LC) droplets configured in either bipolar or radial orientation.[26] In Figs. 3a to 3d, reconstructed principal RI

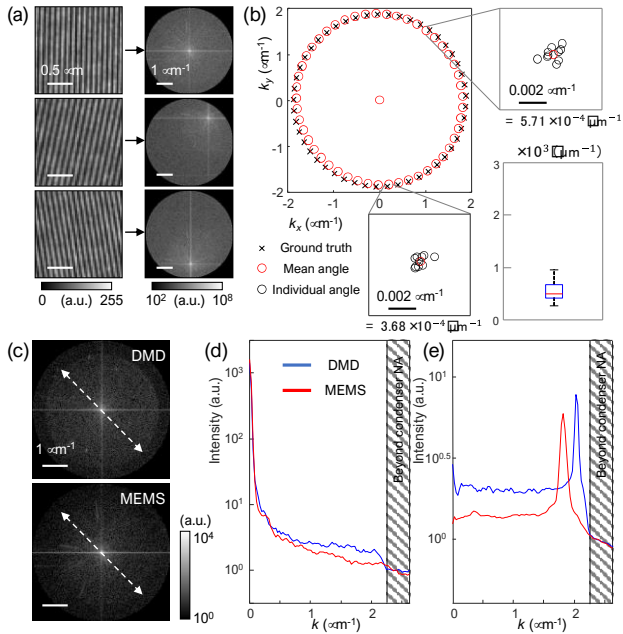

Figure 2. Assessment of MEMS mirror performance and diffraction noise comparison. (a) Interferogram fringe patterns at varied angles and corresponding Fourier space representation without a sample. (b) Circular scanning results

tomograms ($n_1$, $n_2$, and $n_3$), with optic axis orientations, are presented. The orientations respectively agree with the known radial and bipolar structures, demonstrating the effectiveness of both methods in controlling the polarized illumination. Figure 3e and 3f exhibit RI histograms in the background regions of tomograms. The reduced standard deviation in the proposed method indicates a noise reduction compared to the image using DMD.

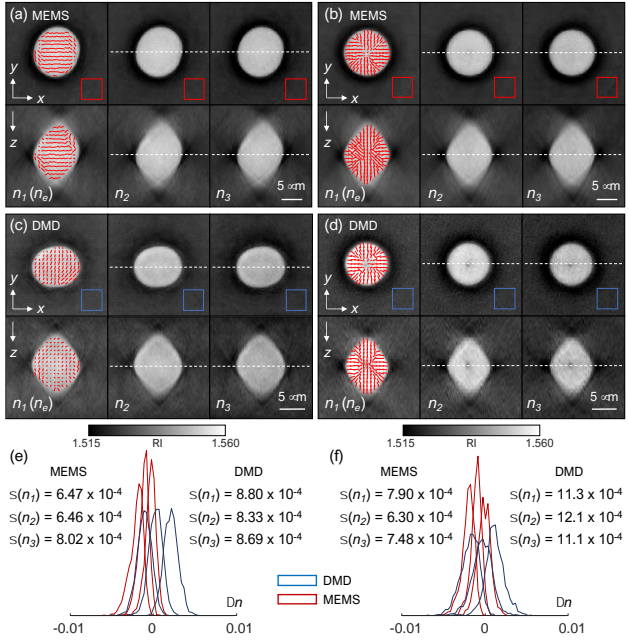

Figure 3. Comparative 3D principal RI tomograms and background noise analysis of liquid crystal droplets. Tomograms reveal bipolar (a, c) and radial (b, d) orientations using MEMS mirror (a, b) and DMD (c, d) methods. Insets show optic axis orientation overlays on n1 slices. Histograms (e, f) compare background noise, with standard deviations indicating noise levels. Background analysis regions are marked in red (MEMS) and blue (DMD) on the tomograms.

We advanced our investigation by applying the MEMS mirror-based DTT method to a multifaceted sample consisting of an array of bipolar liquid crystal droplets. To address the issue of missing cone artifacts typically encountered in such complex reconstructions, we incorporated total variation regularization along with a positive semi-definiteness constraint into our methodology.[27] Figure 4a illustrates the comprehensive 3D distribution of the optic axis directors within the sample. For enhanced clarity, Figure 4b segments the tomogram to distinctly showcase the individual droplets, accentuating their bipolar configurations[28, 29]. The tomograms present an assortment of bipolar droplets, each with unique orientations and dimensions, effectively showcasing the method's robustness in achieving high signal-to-noise ratios.

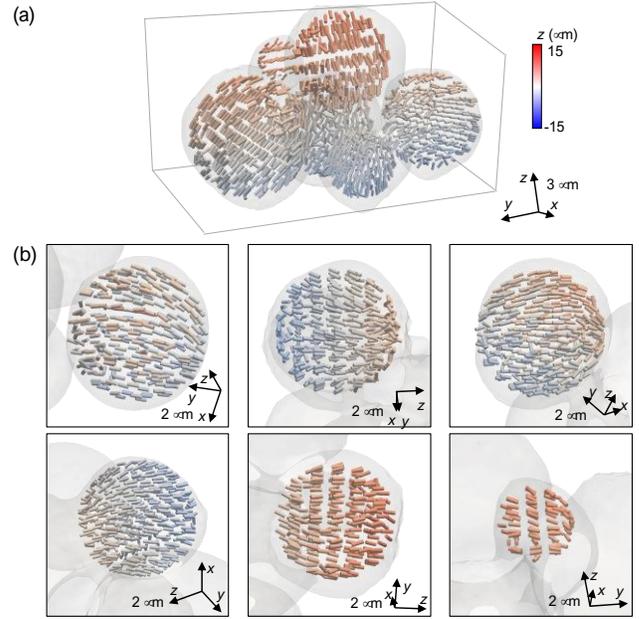

Figure 4. Dielectric tensor tomography of a composite sample with multiple bipolar liquid crystal droplets. (a) An overall 3D rendering of optic axis orientations, highlighted with isosurfaces for $n_e$ = 1.538. (b) A segmented view of the tomogram from (a), detailing individual droplets and their respective bipolar orientations, demonstrating the method's capacity to discern distinct directionalities within a complex structure.

**Conclusion**

In conclusion, this study has established the utility of a MEMS mirror for achieving stable and precise control of illumination angles in DTT. Our comprehensive evaluation of the MEMS mirror's performance, particularly its angle accuracy, precision, and diffraction noise profile, clearly demonstrates its superiority over the DMD. Empirical validation was further obtained by measuring dielectric tensor tomograms of liquid crystal droplets with known anisotropic configurations, which confirmed the MEMS mirror's compatibility with polarized illumination and its precise angle control in DTT. Comparative assessments of background noise in principal refractive index tomograms underscored the high signal-to-noise ratio facilitated by the MEMS mirror, especially in complex samples.

The MEMS mirror brings several key advantages to DTT applications. Its negligible diffraction noise contributes to an enhanced SNR, while its compact design simplifies the optical setup, allowing for rapid and stable imaging processes. Additionally, the MEMS mirror's high diffraction efficiency could enable the use of lower power lasers, reducing costs.

Despite these advancements, the method does encounter limitations, such as speckle noise, associated with the use of highly coherent light sources in off-axis Mach-Zehnder interferometric microscopy. Recent studies suggest that employing low-coherence light sources[30] and use the use of reference-free quantitative phase imaging techniques[31, 32] can significantly improve imaging quality, potentially enhancing DTT. These techniques, along with the unique label-free and quantitative imaging contrasts provided by dielectric tensors, have promising applications, from analyzing topological defects in liquid-crystal materials[33-35], cell mitosis[36, 37], obstetrics and gynecology[38-40], to histopathological evaluations[33, 41, 42]. They offer substantial benefits over traditional polarization microscopy and staining methods, especially in terms of assay time, cost, and the ability to quantitatively assess collagen fiber structures related to tumor metastasis.

Looking to the future, we anticipate that the integration of regularization techniques with DTT could be widely applied, extending possibly to the study of newly emerging active soft materials. The MEMS mirror-based approach, with its notable improvements in image quality and SNR, stands to significantly advance DTT technology and its applications.

## AUTHOR INFORMATION

### Corresponding Author

**YongKeun Park** - *Department of Physics, Korea Advanced Institute of Science and Technology (KAIST), Daejeon 34141, Republic of Korea; KAIST Institute for Health Science and Technology, KAIST, Daejeon 34141, Republic of Korea; Tomocube Inc., Daejeon 34109, Republic of Korea*; Email: yk.park@kaist.ac.kr

### Authors

**Juheon Lee** - *Department of Physics, Korea Advanced Institute of Science and Technology (KAIST), Daejeon 34141, Republic of Korea; KAIST Institute for Health Science and Technology, KAIST, Daejeon 34141, Republic of Korea*

**Byung Gyu Chae** - *Holographic Contents Research Laboratory, Electronics and Telecommunications Research Institute, Daejeon 34129, Republic of Korea*

**Hyuneui Kim** - *Holographic Contents Research Laboratory, Electronics and Telecommunications Research Institute, Daejeon 34129, Republic of Korea*

**MinSung Yoon** - *Holographic Contents Research Laboratory, Electronics and Telecommunications Research Institute, Daejeon 34129, Republic of Korea*

**Herve Hugonnet** - *Department of Physics, Korea Advanced Institute of Science and Technology (KAIST), Daejeon 34141, Republic of Korea; KAIST Institute for Health Science and Technology, KAIST, Daejeon 34141, Republic of Korea*

### Author Contributions

‡J.L. and B.C. contributed equally. The manuscript was written through contributions of all authors. All authors have given approval to the final version of the manuscript.

### Funding

This work was supported by National Research Foundation of Korea (2015R1A3A2066550, 2022M3H4A1A02074314), Institute of Information & communications Technology Planning & Evaluation (IITP; 2021-0-00745) grant funded by the Korea government (MSIT), KAIST Institute of Technology Value Creation, Industry Liaison Center (G-CORE Project) grant funded by MSIT (N11230131).

### Notes

The authors declare no competing financial interest.